\documentclass[fleqn,10pt]{olplainarticle}

\title{Paper scoring through individual reviewer quality estimates for Open Peer Review}

\author[1,2*]{Andrii Zahorodnii}
\author[4]{Jasper J.F. van den Bosch}
\author[5]{Ian Charest}
\author[6]{Christopher Summerfield}
\author[2,3*]{Ila R. Fiete}
\affil[1]{Department of Electrical Engineering and Computer Science, MIT}
\affil[2]{Department of Brain and Cognitive Sciences \& McGovern Institute, MIT}
\affil[3]{K. Lisa Yang Integrative Computational Neuroscience Center, MIT}
\affil[3]{School of Psychology, University of Leeds, United Kingdom}
\affil[4]{Département de Psychologie, Université de Montréal, Canada}
\affil[5]{Department of Experimental Psychology, University of Oxford, United Kingdom}
\affil[*]{Correspondence: zaho@mit.edu (AZ), fiete@mit.edu (IRF)}


\begin{abstract}
Traditional closed peer review systems, which have played a central role in scientific publishing, are often slow, costly, non-transparent, stochastic, and possibly subject to biases -- factors that can impede scientific progress and undermine public trust. Here, we propose and examine the efficacy and accuracy of an alternative form of scientific peer review: through an open, bottom-up process. First, using data from two major scientific conferences (CCN2023 and ICLR2023), we highlight how high variability of review scores and low correlation across reviewers presents a challenge for collective review. We quantify reviewer agreement with community consensus scores and use this as a reviewer quality estimator, showing that surprisingly, reviewer quality scores are not correlated with authorship quality. Instead, we reveal an inverted U-shape relationship, where authors with intermediate paper scores are the best reviewers. We assess empirical Bayesian methods to estimate paper quality based on different assessments of individual reviewer reliability. We show how under a one-shot review-then-score scenario, both in our models and on real peer review data, a Bayesian measure significantly improves paper quality assessments relative to simple averaging. We then consider an ongoing model of publishing, reviewing, and scoring, with reviewers scoring not only papers but also other reviewers. We show that user-generated reviewer ratings can yield robust and high-quality paper scoring even when unreliable (but unbiased) reviewers dominate. Finally, we outline incentive structures to recognize high-quality reviewers and encourage broader reviewing coverage of submitted papers. These findings suggest that a self-selecting open peer review process is potentially scalable, reliable, and equitable with the possibility of enhancing the speed, fairness, and transparency of the peer review process.
\end{abstract}

\begin{document}

\flushbottom
\maketitle
\thispagestyle{empty}

\section{Introduction}
Modern technologies have empowered the sharing of information at scale, as well as commentary and feedback on the shared content. However, distilling this collective feedback into a reliable collective assessment of shared information has remained a thorny challenge. The well-known and ubiquitous problem of mis- and dis-information on social media platforms is a testament to this difficulty \citep{vosoughiSpreadTrueFalse2018, Kitchens2020-fl}.  

Scientific communication, the primary focus of the present work, suffers from related problems. Partially for historical reasons \citep{Birukou2011-wu} and partly to head off the kinds of misinformation problems found in open media, scientific review is a closed, slow, top-down, expensive process with incentives for high-profile publishing that might also distort science \citep{Van_Noorden_2013, Young_2008, Guardian-history-17, Teixeira_da_Silva_2014}. A scientific report is submitted to one journal, whose editors select whether to desk-reject it or send it out for review to a small number of hand-picked reviewers. The resulting reviews emanate from a small number of reviewers and are thus highly stochastic. If the paper is rejected, the process repeats until the paper is eventually accepted at some journal. This friction-filled process slows scientific progress in a way that can impede junior scientists' careers. The exponentially growing volume of scientific output makes these challenges more daunting. The concept of "publish then review", with bottom-up review, offers a potential antidote \citep{eisenImplementingPublishThen2020, Ginsparg_1997, Eisen16, Kravitz_2011, Kriegeskorte_2012, Nosek_2012, Stern_2019, Teixeira_da_Silva_2013, 10.3389/fncom.2012.00032}. Submitted papers are immediately considered as published, and review is an ongoing process that begins after publication. This process requires making reviewing open, equitable, and bottom-up, so that any scientist can 
review any paper by self-selection \citep{LeCun13}. However, how to make such an open system reliable, and with the right incentive mechanisms, is an open problem \citep{Sandewall2012-be, Birukou2011-wu, Poschl2012-wk, goldberg2024peerreviewspeerreviews}. 

In this paper, we take a data-driven approach to the question of how to extract a reliable signal from noisy and flawed data provided by the scientific community. We first quantify the level of agreement between reviewers in peer review data, and show that it is surprisingly small. Next, we propose a method for assessing paper quality based on an estimate of reviewer quality for each individual. Interestingly, reviewership quality is independent of authorship quality. We show that this metric can support an open, democratic peer-review solution that tackles the following challenges: it allows reviewers to self-organize by picking the works that they want to review; it provides recognition and thus incentives to reviewers for consistently writing high-quality reviews; in addition, it allows an effective extraction of the community belief about the quality of papers from scarce and noisy reviews.

\section{Results}
\subsection{High variability: A core challenge for collective paper review}

\begin{figure}
\centering
\includegraphics[width=0.8\linewidth]{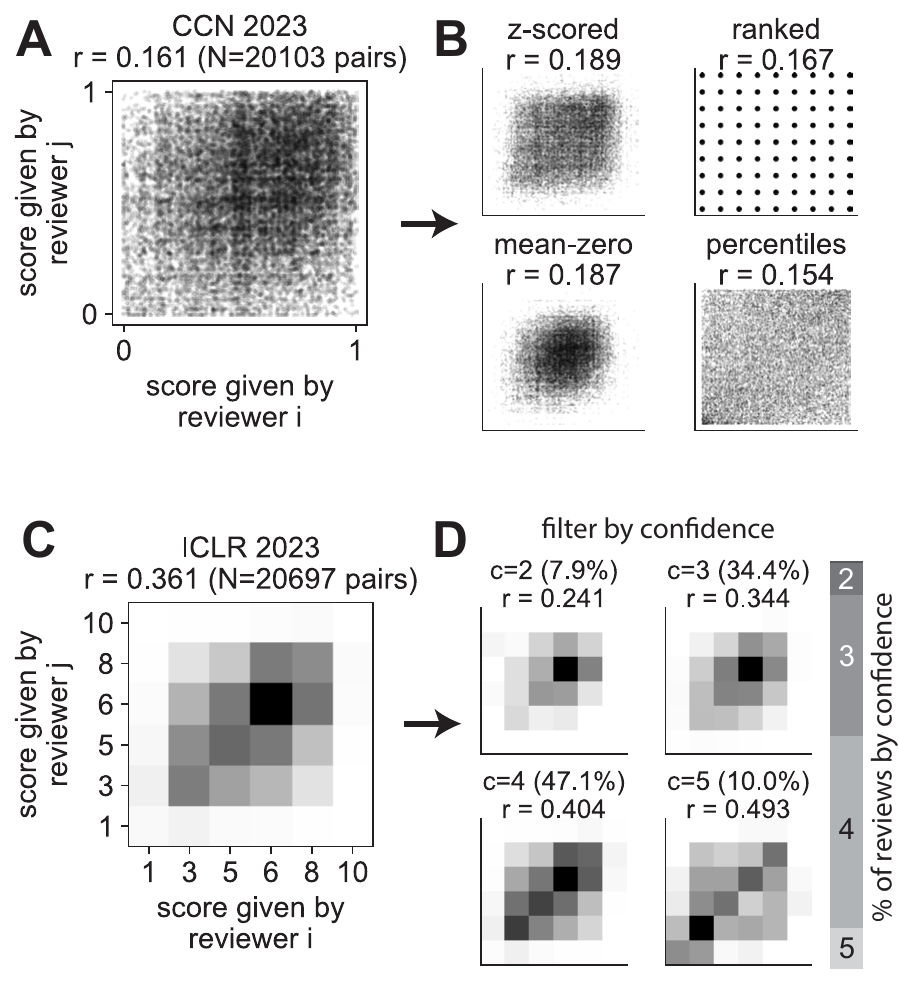}
\caption{{\bf High variability in paper scores and low correlation across reviewers are challenges for collective paper review.} 
(A) Correlation of scores between pairs of reviewers reviewing the same submission in the conference Cognitive Computational Neuroscience 2023 ("Impact" scores). 
(B) Common score-normalizing techniques -- z-scoring, ranking, removing the mean, and inverting the distribution, applied to the CCN2023 data: the correlations remain low.
(C) Corresponding to (A) analysis for the International Conference on Learning Representations 2023.
(D) In ICLR2023 data, pairs of ratings with higher self-reported confidence scores have a higher correlation, but only a minority of reviews is tagged as high-confidence.
For all panels showing correlation, $p<0.001$.}
\label{fig:1}
\end{figure}

We consider data from two conferences that employed a transparent review process and used explicit scoring for determining paper acceptance decisions: the International Conference on Learning Representations (ICLR) 2023 and the Conference on Cognitive Computational Neuroscience (CCN) 2023. In the CCN2023 peer review process, each of the 589 reviewers assigned "Impact" and "Clarity" numerical scores to the 527 submitted abstracts. All reviewers were authors of submitted abstracts. Each abstract received an average of 9.1 reviews. In ICLR2023, each of the 3798 submitted papers received anywhere from 2 to 9 reviews by reviewers selected through a separate track, with an average of 3.8 reviews per submission. Each review consisted of multiple dimensions of scores, with the overall rating and confidence scores used in analyses in this paper.

In both datasets, the correlation in scores given by pairs of reviewers to the same paper was surprisingly low (Figure \ref{fig:1}). The correlation was very weak in the community-based review process of CCN ($r=0.161\pm0.014$; Figure \ref{fig:1}A). Though somewhat higher for ICLR2023 ($r=0.361\pm0.014$; Figure \ref{fig:1}C), the correlations were still small despite the fact that the submissions being assessed were full-length standalone papers. Common possible causes of low correlation, such as variations in what range of scores each reviewer actually utilizes, are sometimes addressed by score normalization techniques (e.g., z-scoring, ranking, mean removal, or distribution inversion). Applying any of these methods only marginally increased the correlation ($r<0.19$ for CCN2023; Figure \ref{fig:1}B). Although the correlation increased with the self-reported confidence scores in ICLR2023 data (from $r=0.24$ for confidence=2 up to $r=0.49$ for confidence=5; Figure \ref{fig:1}D), only $10.0\%$ of reviewers reported having this high confidence. Moreover, the confidence of ratings was inversely correlated with the paper quality rating scores (Supplementary Figure \ref{figsupp:sf1}), implying that on average, better papers receive less confident reviews. In sum, our analysis reveals a striking lack of agreement among reviewers when assessing the same paper.

\subsection{Bayesian paper quality estimation using empirical reviewer quality estimates}

\begin{figure}
\centering
\includegraphics[width=0.8\linewidth]{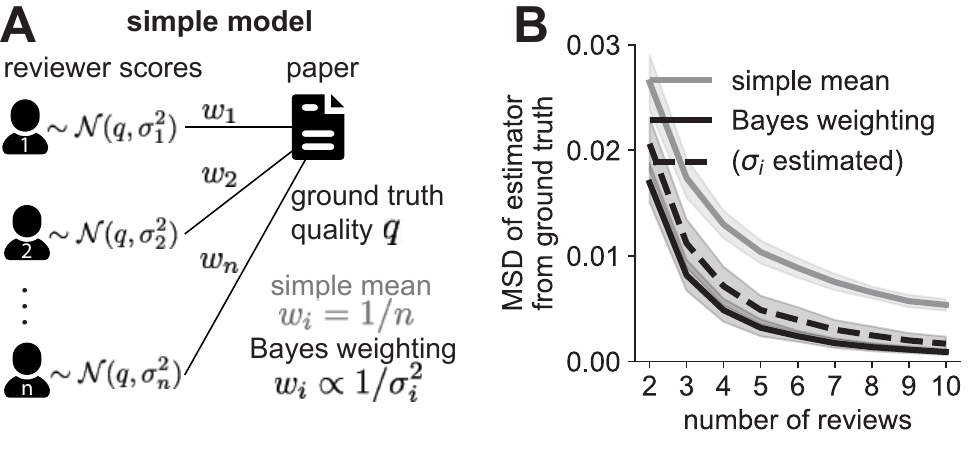}
\caption{{\bf In a simple model, weighting reviews using an empirical reviewer quality metric results in better estimation of paper quality.} (A) The simple model: different reviewers all assign scores to the same paper according to a distribution centered at the ground truth quality (hidden), but with different standard deviations, corresponding to different reviewer accuracies. 
(B) Bayesian weighting of review scores based on  reviewer quality leads to a tighter estimation of the ground truth paper quality scores. Curves shown for ground truth reviewer quality  (black solid line), empirically estimated reviewer quality based on five previous reviews (black dashed line), and simple mean score (solid gray line). Shaded regions denote the s.d.}
\label{fig:2}
\end{figure}

\paragraph{Methods: Estimating paper quality using reviews.}
To demonstrate how a potential remedy to this challenge can look like, we first consider a very simple model (Figure \ref{fig:2}; a slightly richer model is presented below). Assume that each submission (paper) $j$ has a hidden ground truth quality $q_j$. Reviewers are assumed to give scores to that paper drawn from a distribution centered at $q_j$, with a user-specific standard deviation $\sigma_i$, where $i$ is the index of the reviewer (Figure \ref{fig:2}A). We can think of the inverse standard deviation of a reviewer as a ground-truth reviewer quality measure. The goal is to obtain good estimates of $q_j$ without knowledge of the paper ground truth values. This simple model leaves out possible systematic reviewer biases (see \cite{neurips-process-16,Helmer_2017,Ross_2017, double-blind-19, goldberg2024peerreviewspeerreviews, stelmakh2023citereview} for discussion of biases in the review process), assumes there are no bots or bad actors (we add bots below), and doesn't constrain the scores to lie in a bounded interval, 
however it suffices to demonstrate the core idea. 
Traditionally, the scores of all reviewers are averaged to obtain a final estimated quality score of the paper, the score on which decisions are made. This calculation produces an estimator of the paper quality score with mean squared deviation (MSD) equal to
\begin{equation}
    \text{MSD(simple mean)}=\frac{\sum \sigma_i^2}{n^2}.
\end{equation}
 
However, if reviewer quality is known, the optimal weighting technique (a Bayesian approach) is to assign weights to reviewers' scores that are proportional to the inverse square of the estimated standard deviation of that reviewer (for derivation of these results, see Appendix 1): $w_i\propto 1/\sigma_i^2$. This approach leads to a paper quality estimator with a strictly lower MSD (unless all $\sigma_i$ are the same):
\begin{equation}
    \text{MSD(Bayesian)}=\frac{1}{\sum 1/\sigma_i^2}.
    \label{cert_est}
\end{equation}
This Bayesian paper quality estimator based on known reviewer quality follows the ground truth quality of the paper much more tightly (Figure \ref{fig:2}B, solid black) than the standard mean  (Figure \ref{fig:2}B, gray).

In real settings, the reviewers' quality is of course not known {a priori}, requiring that this parameter be estimated from the data. To simulate this situation, we compute an empirically estimated reviewer quality based on five previous reviews by each reviewer, by using the mean squared deviation of those reviews from the community average paper score. In this process, we assume that the community score was collected from enough other reviewers to estimate the ground truth qualities of those papers with high accuracy and no bias.
When we apply Bayesian paper quality estimation with the empirically determined reviewer quality estimates, we find that the strategy still significantly outperforms the standard mean (Figure \ref{fig:2}B, black dashed line). Our simple model setting, therefore, suggests that estimating individual reviewer quality and applying those quality estimates in a Bayesian way to papers can can improve estimates of paper quality. Next, we test these ideas in on real-world conference data.

\subsection{Applying the Bayesian approach to CCN 2023 review data}

\begin{figure}
\centering
\includegraphics[width=0.8\linewidth]{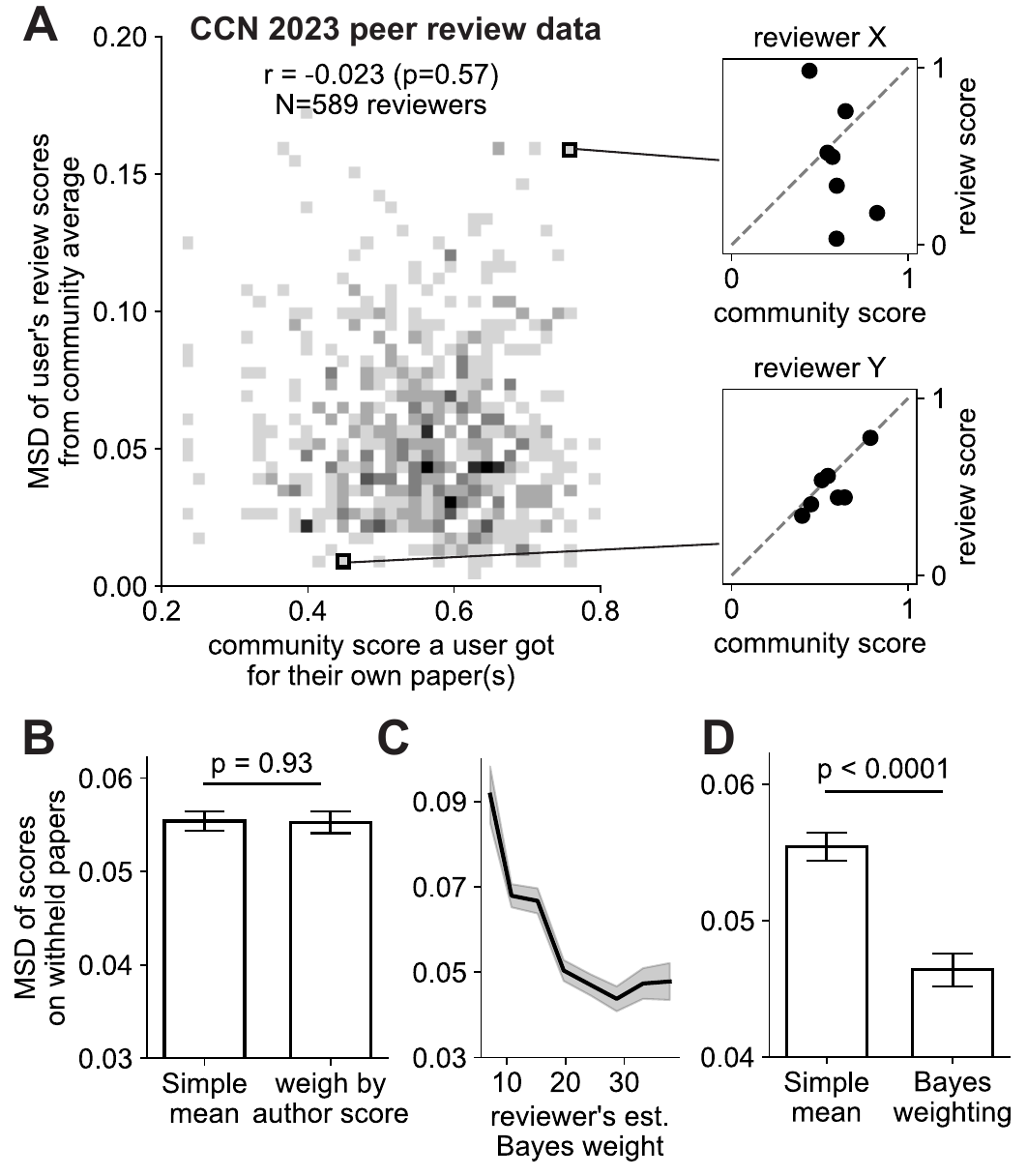}
\caption{{\bf Reviewer quality, unrelated to author quality, improved paper quality estimation in CCN2023 submissions.} 
(A) In the CCN2023 review process, reviewer quality (measured as MSD of user's review scores from community average score across all reviewed papers) and author quality (measured as the average score the user got for their paper) were unrelated. To preserve reviewer privacy in this visualization, some outlier points were removed from the plot and reviewers were binned together in small bins.
(B) Using estimated author quality as a stand-in for reviewer quality as the Bayes weight did not improve the estimate of paper quality relative to baseline (cross-validated results, shown on withheld papers). Error bars denote the s.e.m.
(C) Users with higher estimated reviewer quality (Bayes weight estimated on the training data) consistently had smaller deviations from community score on withheld papers when estimating reviewer quality. Shaded regions denote the s.e.m. 
(D) Using the estimated reviewer quality as Bayes weights led to consistently smaller deviations from the community score on withheld papers. Error bars denote the s.e.m.}
\label{fig:3}
\end{figure}

To assess outcomes in real-world settings, we turn to the data from CCN2023 (Figure \ref{fig:3}). CCN2023 data contained unique reviewer IDs for each reviewer, across papers, and a unique author ID for each author across their submitted papers. Thus, we were able to consider different methods to compute reviewer quality scores and determine the efficacy of using those scores in paper quality estimation.  

\paragraph{Methods: Proxy for ground-truth paper quality.} The first challenge in using real world datasets is that we lack ground-truth knowledge of paper quality, hampering our ability to compare different estimators of paper quality. Under the assumption that if a large fraction of the community were to assess a paper, the average estimate would be the ground truth measure of quality, we construct a proxy measure for ground-truth: the community average score (CAS) for each paper. There are various important discussions about the true quality of a scientific paper, such as the durability of its findings over time; however, even large-scale expert review cannot estimate durability because it involves complex future prediction. Thus our focus, thoughout this paper, is on the question of how one can arrive at a reasonable estimate of the CAS from a severely limited number of noisy reviews. 

Below, we compute the CAS for each paper as the average of the ``impact" score assigned by all reviewers of that paper. We then assess the efficacy of a paper quality estimation technique by applying it to subsampled sets of reviews for the paper, and comparing the estimate to the CAS. 

\paragraph{Is authorship quality predictive of reviewer quality?} 
Scientists in CCN2023 play two roles, unlike in our simple model: they are both reviewers and authors. A quality value can be assigned to each role. A natural question is whether the two properties interact: for instance, is being a good author predictive of being a good reviewer? If they are related, then given the sparsity of the data (each individual writes relatively few reviews and each paper is reviewed by relatively few reviewers), combining estimates for quality in both roles might enable a more robust and better reviewer quality estimate.

\begin{figure}[h]
\centering
\renewcommand{\figurename}{Supplementary Figure}
\renewcommand{\thepage}{S\arabic{page}}
\renewcommand{\thesection}{S\arabic{section}}
\renewcommand{\thetable}{S\arabic{table}}
\renewcommand{\thefigure}{S\arabic{figure}}
\includegraphics[width=0.9\linewidth]{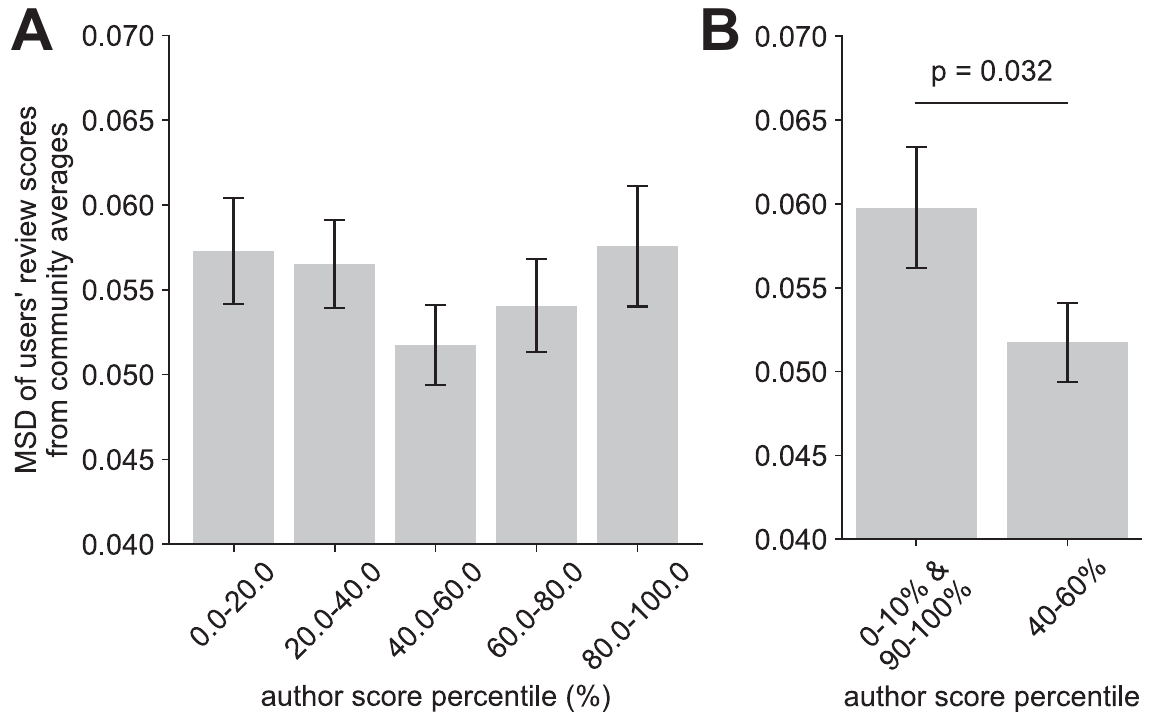}
\caption{Reviewers with intermediate author scores are the best at predicting the community average paper quality score. (A) Distribution of mean standard deviations from community average score across the different reviewer groups, grouped by the authorship score of the reviewer. (B) Reviewers with the author scores in the range between 40th and 60th percentiles have significantly lower mean squared deviation from the community average score than the reviewers from the ends of the spectrum of author scores (0-10th and 90-100th percentiles; $*p < 0.05$). XXX indicate which dataset}\label{figsupp:sf2}
\end{figure}

\begin{figure}[h]
\centering
\renewcommand{\figurename}{Supplementary Figure}
\renewcommand{\thepage}{S\arabic{page}}
\renewcommand{\thesection}{S\arabic{section}}
\renewcommand{\thetable}{S\arabic{table}}
\renewcommand{\thefigure}{S\arabic{figure}}
\includegraphics[width=0.9\linewidth]{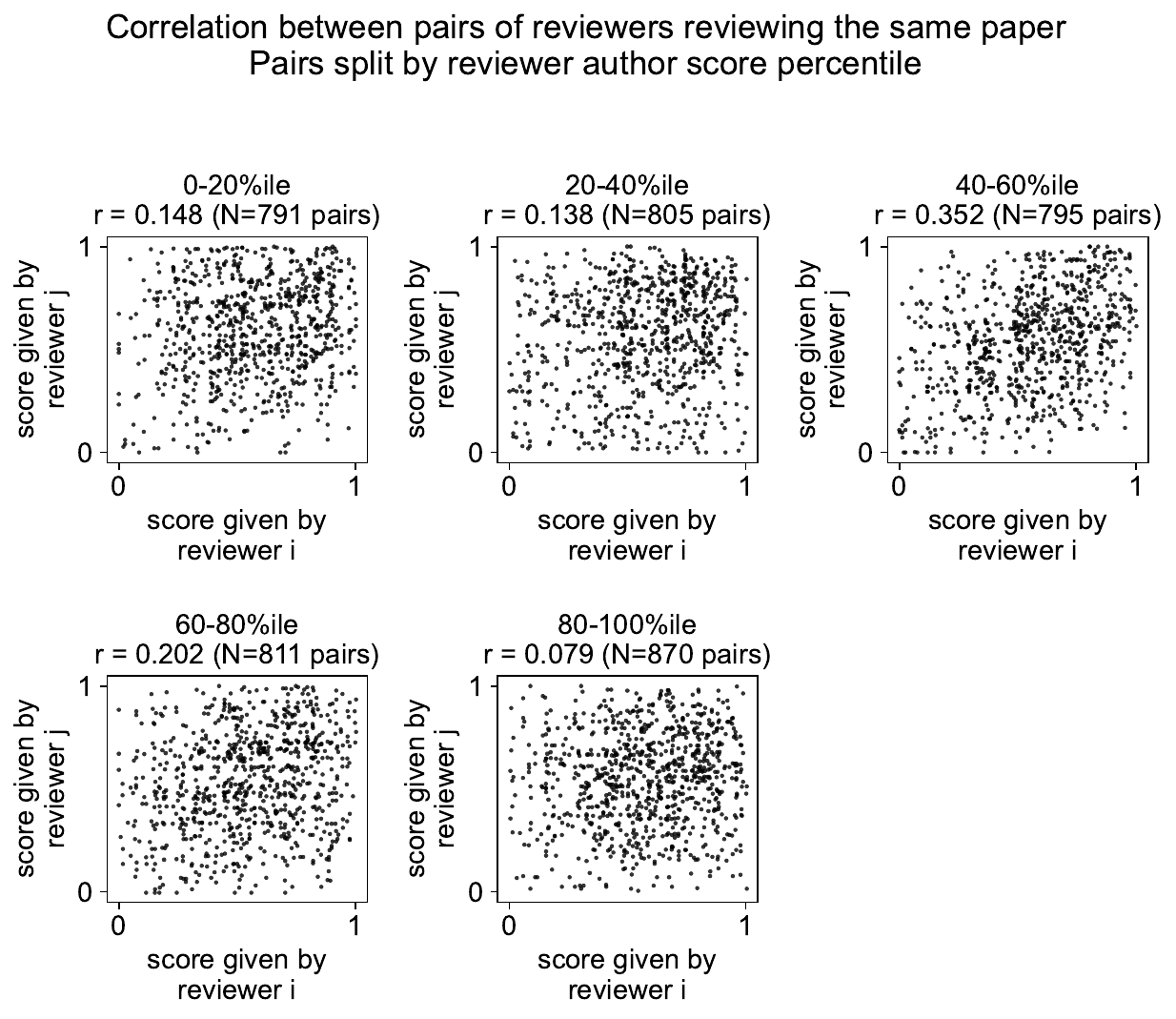}
\caption{Reviewers with intermediate author scores correlate the most with other reviewers within their group. Correlation between pairs of reviewers reviewing the same paper, split by the authorship score of the reviewers. The correlation is the highest for the reviewers with intermediate authorship scores (in the range between 40th and 60th percentiles; $N>90$ pairs, all panels). XXX indicate which dataset}\label{figsupp:sf3}
\end{figure}

To answer this question, we considered the mean squared deviation of an individual's scores from the CAS for each paper reviewed by that individual (i.e., their quality as a reviewer), and compared it with the CAS  they received for their own submissions (i.e., their quality as an author). This resulted in the surprising discovery that authorship quality is unrelated to reviewership quality (Figure \ref{fig:3}A; $r=-0.023, p=0.57, N=539$ reviewers). This finding challenges the assumption that high-scoring authors are necessarily good at predicting the community judgment of a paper's quality. Instead, there were reviewers who received very high scores for their own submissions yet were remarkably poor at predicting the community judgment of papers they reviewed (Figure \ref{fig:3}A, inset, upper right). Conversely, some reviewers who received much lower scores for their own submissions were able to estimate the CAS quality of reviewed papers much more accurately (Figure \ref{fig:3}A, inset, lower right). In fact, the group of reviewers that was able to estimate the CAS with the highest accuracy was the group with intermediate authorship scores (Figure \ref{fig:3}, Supplementary Figure \ref{figsupp:sf2}A-B), and this group was also had the highest level of agreement between reviewers reviewing the same paper (Figure \ref{fig:3}, Supplementary Figure \ref{figsupp:sf3}; $r=0.352$ for the middle percentiles, N=795 pairs). This discrepancy between the authorship score and the reviewer quality might be attributable to various factors, including a lack of rewards and recognition for putting effort into conducting a thorough review. Thus, at least with reviews conducted as they were in CCN2023, reviewer quality appears to be independent of author quality, and author quality likely cannot be used in a direct way as a measure of reviewer quality. 

We test this hypothesis as follows. The baseline method for paper quality estimation as a function of number of reviewers is simple averaging of subsets of reviewer scores. We compare this baseline against a method which weighs the reviewer's scores using the authorship quality of each reviewer, using empirically estimated author quality as a measure of reviewer quality. Authorship quality is estimated by the average of the CAS on all papers of that author. 
Using this empirical authorship quality measure as weights led to no measurable improvement in the MSD around the CAS of the resulting paper quality estimates relative to simple mean estimators, Figure \ref{fig:3}B ($MSD=0.0553 \pm 0.001$ versus $MSD=0.0554 \pm 0.001$ for the mean estimator). 
Thus, we conclude that (estimated) author quality is not directly proportional to (estimated) reviewer quality and does not, applied in a simple way, result in better paper quality estimation. 

In the results of this section, the cross-validation was done in a leave-one-out fashion: that is, to estimate the deviation of a reviewer's score of a paper from the CAS, that reviewer was temporarily removed from the dataset, the CAS was calculated without it, and then compared to that reviewer's score to obtain the deviation.

\paragraph{Improved CCN 2023 assessments through empirical Bayesian reviewer quality weighting}

We next consider whether and how direct reviewer quality estimates can improve paper quality estimation, according to the procedures of our simple model above (Figure \ref{fig:2}). To do so, for every paper in the dataset, we estimate its reviewers' Bayes weights by leaving this paper out of the dataset, then calculating the mean squared deviation of the reviewers' scores from CAS on all the other papers, and using the formula $1/\sigma_{\text{estimated}}^2$ to obtain the Bayes weights for each reviewer according to Eq. \ref{cert_est}.
We were able to perform these tests in the CCN 2023 dataset but not the ICLR data because ICLR's process does not provide a consistent reviewer identifier across reviewed papers. 

To obtain a cross-validated result that avoids information leakage between the data used to obtain reviewer quality scores and the data on which we evaluate paper quality, we only consider performance on withheld papers -- those not included in the estimation of the users' reviewer quality scores. We find that indeed, individuals with a higher estimated reviewer quality score consistently demonstrated a better cross-validated ability to predict the CAS score (Figure~\ref{fig:3}C). 

Given this result, we next generate a cross-validated prediction of the CAS by combining paper scores from reviewers based on their reviewer quality scores. We find that this procedure yields a better prediction of the CAS than simple averaging of paper scores (Figure \ref{fig:3}D): performance is $MSD=0.046 \pm 0.001$, compared to a higher $MSD=0.055 \pm 0.001$ with the simple mean estimator ($p<0.001$). 

In sum, this estimation process -- which first estimates reviewer reliability based on a subset of their reviews and the CAS, then performs Bayesian weighting of paper scores -- has the potential to enhance the overall quality and fairness of peer reviews when the number of reviews per paper is limited (in this dataset, the mean number of reviews per paper is 9). Next, we consider an alternative method for estimating reviewer quality, based on direct peer evaluation of reviews, and compare the two methods. 

\subsection{Open peer review with user-generated ratings of both papers and reviews}

\begin{figure}[h!]
\centering
\includegraphics[width=\linewidth]{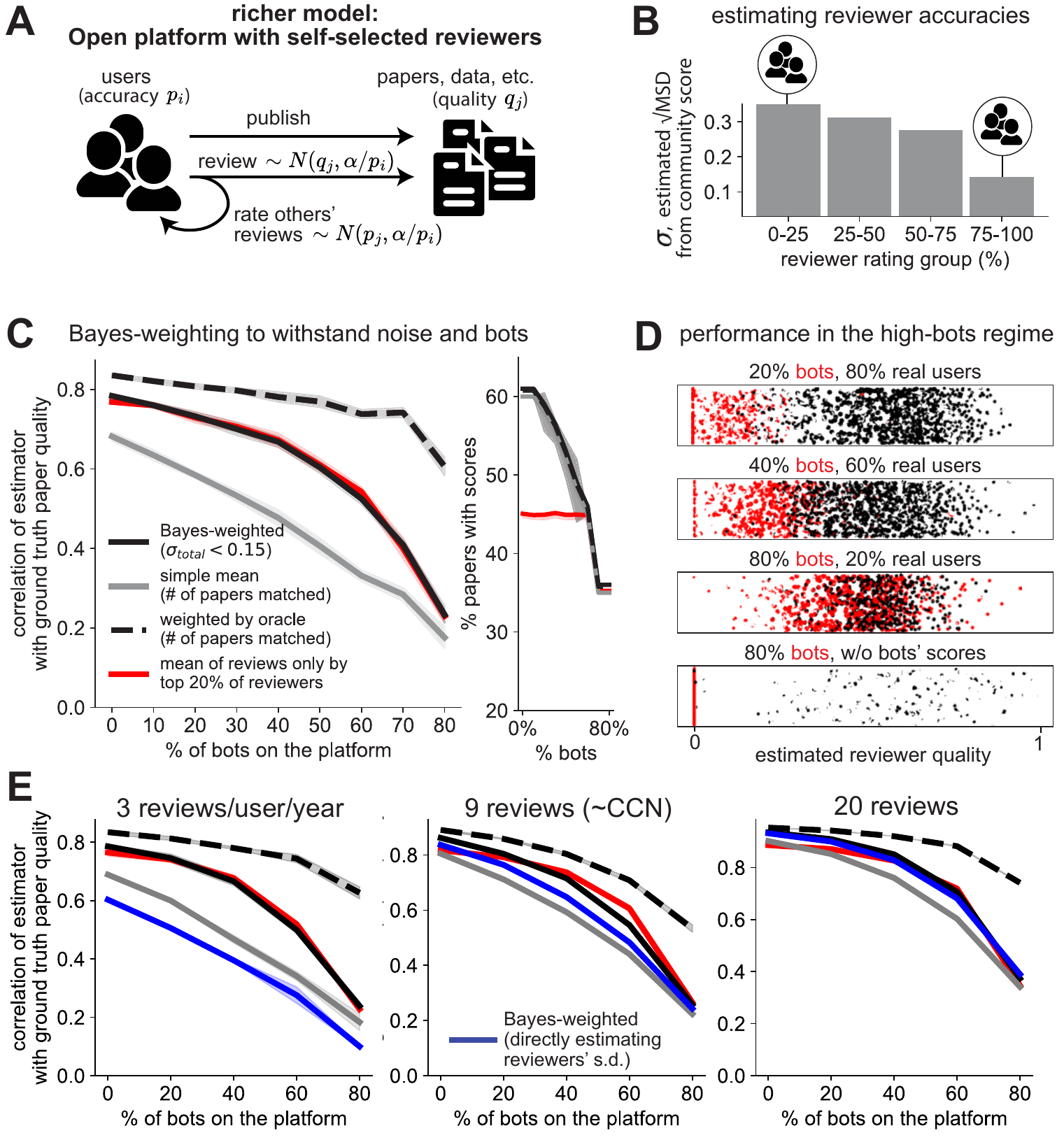}
\caption{{\bf Reviewing reviews: Benefits of reviewer quality metrics in a richer ecosystem.} 
(A) An open platform where content is published immediately, reviewers are self-selected, and the community assesses (rates) review quality. 
(B) Estimation of reviewer accuracy: binning reviewers by the ratings their reviews received enables estimation of that group's accuracy in paper evaluation (mitigating the problem of estimating an accuracy score based on individual reviewer scores, given the small number of reviews supplied by each reviewer). 
(C) Bayes weighting of review scores based on reviewer accuracy as estimated from (B) leads to a better estimate of the ground truth paper quality scores (left; solid black line) than simple averaging of all review scores (left; gray line). An alternative heuristic of simply thresholding out all reviewers with a sufficiently low rating percentile (here, below 80 percentile, red line) results in a similar correlation to Bayes weighting but much lower coverage of papers on the platform (right).
(D) Distributions of estimated reviewer quality scores for bots and real users in the simulated platform. In high-noise regime with most of the users being bots (up to 80\%), reliability of the reviewer quality estimation breaks down, suggesting that alternative ways of excluding bots might be more effective.}
\label{fig:4}
\end{figure}

Building on the findings above, we hypothesize that a new open framework for scientific peer review may be feasible and scalable, involving bottom-up self-selection of papers to review by users, and users scoring others' reviews  (Figure \ref{fig:4}).

In the hypothesized framework, all papers are immediately published \citep{eisenImplementingPublishThen2020}. Post-publication, users on the platform self-select themselves to review the paper. Users may give quality ratings not only to papers but also to others' reviews \citep{10.3389/fncom.2012.00032}. The quality of the given publication is estimated from these reviews and the ratings of the reviews. Authors of new submissions, reviewers, and raters of reviewers all come from the same user pool. In the subsections below, we assess the reliability of this process. 

\paragraph{Methods: Generative model} We implement a generative model of this process (Figure \ref{fig:4}A): 
The model consists of a set of papers and a set of reviewers. Each generated paper $i$ has a hidden ground truth quality $q_i\in(0, 1)$. Each reviewer has a ground truth reviewer quality $p_i\in(0, 1)$. Whenever reviewer $i$ writes a review for paper $j$, they assign to it a score $\sim N(q_j, \alpha/p_i)$: in other words, the score is based on the intrinsic paper quality, with a standard deviation that is inversely proportional to reviewer quality. Whenever user $i$ rates a review of another user $j$, they assign a score to them which is $\sim N(p_j, \alpha/p_i)$. We set $\alpha = 0.18$ to match the reviewer correlation levels to those we found in CCN2023 data. We set the standard deviations of the review scores and paper scores to be equal, based on the finding from a NeurIPS 2022 randomized control trial \citep{goldberg2024peerreviewspeerreviews} that disagreement between reviewers assessing review quality are comparable to the disagreement rates of paper reviewers. In our experiments, arying $\alpha$ did not change the results qualitatively. In addition, we consider the existence of consistently low-quality, unreliable reviewers or ``bots''. These unreliable users are modeled as assigning paper review scores and ratings of other reviews uniformly at random in $[0,1]$, regardless of paper or review quality.

All assigned scores must lie in the interval $[0, 1]$, and user-assigned scores are the only data used by the system to derive its quality estimators. As a conservative scenario, we first consider a batch/offline process, in which users have provided reviews of papers and other users, and we must estimate all quantities (paper quality and reviewer quality) at the same time: scoring is based on all available papers and reviews available at the end of the review process (a history-dependent process, which we discuss below, may enable further accuracy gains). We further assume, as a conservative poor-case scenario and consistent with the CCN 2023 dataset, that we are in a regime where assessing quality is hardest: the low-review regime in which each user publishes one paper a year, reviews $\leq$ 3 papers and rates $\leq$ 10 other users, meaning that each paper receives 3 reviews on average.

\paragraph{Methods: Scoring} Scoring proceeds as follows: a rating is calculated for each reviewer by averaging all scores that this reviewer received for their reviews from other users. We then bin reviewers by their reviewer rating, and estimate how closely each bin predicts the CAS for any submission (Figure \ref{fig:4}B). We also estimate the mean squared deviation (MSD) of each bin's scores around the CAS, on the subset of papers on the platform for which the CAS is known with high certainty (in this example we use the papers with most number of reviews received -- top 20\% -- when only including reviews by reviewers who have a rating in the top 20\% percentile). 

The MSD of the bin is used as the MSD for each reviewer in the bin. The binning procedure helps to mitigate the problem that each reviewer generally scores very few papers and each paper generally receives very few reviews, for accurate assessment of MSD. In addition, the binning removes the incentive to copy other reviewers' assessments, preventing reviewers from artificially inflating the measure of others' agreement with their scores. With this estimated MSD for each reviewer, we can now use Bayes weighting, as in the simple model of Figure \ref{fig:2}, to generate a quality score for each article on the platform based on the reviews received.

Computing the MSD of reviewers allows for the generation of an estimation certainty score for each paper, based on the equation $1/\sigma^2_\text{total}=\sum 1/\sigma_i^2$ (Appendix 1; Eq. \ref{cert_est}). We combine this weighting procedure with committing to publish only those scores that pass a certain threshold of certainty (here -- $\sigma_\text{total}<0.15$). 
  
\paragraph{Direct user ratings of reviews improves paper scoring}
The result of the procedure above is an estimator of paper quality (Figure \ref{fig:4}C; solid black line) that tracks the (hidden) ground truth quality of papers on the platform more closely than the simple mean estimator (Figure \ref{fig:4}C; solid gray line). It is possible to create an estimator of paper quality through a procedure that interpolates between a simple mean and the Bayesian estimator, by considering and averaging together (without further weighting) only the reviews submitted by the top 20\% of reviewers for each paper. The performance of this estimator on a given paper matches that of the Bayesian estimator (Figure \ref{fig:4}C; solid red line), but has the downside that its coverage of papers is sharply reduced, with assessments available for a much smaller number of papers (only 45\% compared to 60\% of papers with the Bayesian method; Figure \ref{fig:4}C, right). Thus, our Bayesian estimator makes more efficient use of existing reviewers, including poorer reviewers, to arrive at good quality assessments of a larger fraction of papers.

We compared the performance of the Bayesian estimator above, which is based on direct user evaluation of review quality with Bayesian estimation performance, with the simpler reviewer quality estimation procedure (from Figure~\ref{fig:2}B and the CCN data analysis of Figure~\ref{fig:3}), which ignores the scores given by users to reviews and estimates each reviewer's S.D. directly based on the deviations of their reviews from the CAS. The performance of this procedure depends heavily on the mean number of reviews each paper receives: in a typical situation of 3 reviews per paper, it performs even worse than simply averaging of all review scores (Figure~\ref{fig:4}E, left). When the number of reviews approaches that of CCN data (around 9), performance is slightly better than the simple mean (Figure~\ref{fig:4}E, middle), consistent with our finding in real CCN data (Figure~\ref{fig:3}). As the number of reviews grows to 20, the performance of the direct S.D. estimator matches that of our best proposed measure from the above subsection (Figure~\ref{fig:4}E, right). However, there are two serious drawbacks: this procedure should not naively be implemented in any online peer review platform, as it rewards reviewers who have small score deviations from CAS, encouraging reviewers to give papers the same score as all the other reviewers. This incentive can lead to a collapse scenario where each reviewer simply selects whatever rating was popular among other reviewers. And, even post-factum, this method only matches the direct user evaluation of reviewer-quality when the number of reviews is large.

\paragraph{Reducing unreliable reviewers}
Unreliable reviewers, defined in our model as users who assign random reviews that are uncorrelated with user and paper quality, may map in reality to different populations: they could be scientists who invest minimal time or effort in the review process, or they could be adversarial human or even non-human (''bot'') actors. 

\begin{figure}[h]
\centering
\renewcommand{\figurename}{Supplementary Figure}
\renewcommand{\thepage}{S\arabic{page}}
\renewcommand{\thesection}{S\arabic{section}}
\renewcommand{\thetable}{S\arabic{table}}
\renewcommand{\thefigure}{S\arabic{figure}}
\includegraphics[width=0.5\linewidth]{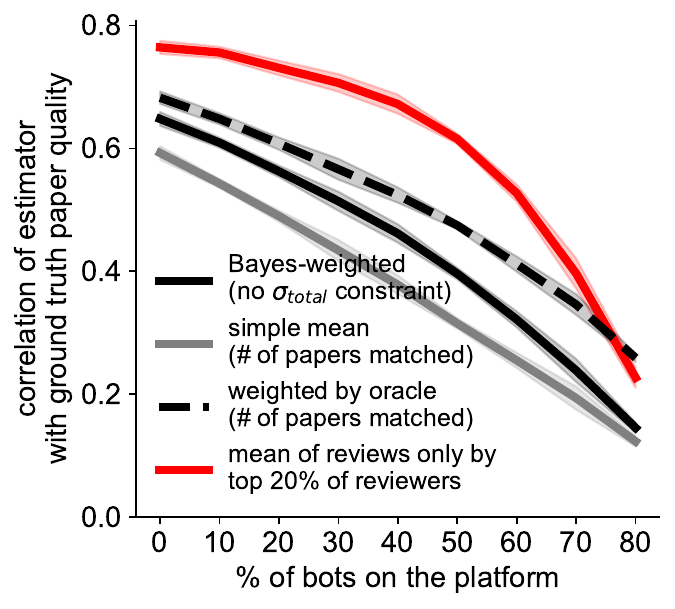}
\includegraphics[width=0.2\linewidth]{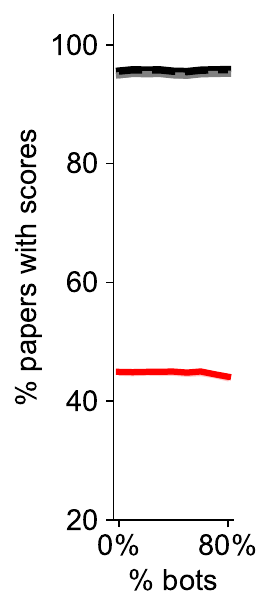}
\caption{Including a minimum confidence threshold (an upper bound of $\sigma_{\text{total}}$) is necessary, for without it, the correlation between true and estimated paper quality drops, even when using the oracle to estimate the scores.}\label{figsupp:sf4}
\end{figure}

One way to reduce the influence of non-human bots is to require all users who deposit papers or review papers to be verified scientists \citep{10.3389/fncom.2012.00032}. Such a step is essential because the number of non-human bots can overwhelm the number of human users by orders of magnitude. After this step, in the low-review regime with the possibility of unreliable human users, imposing a constraint on minimum certainty about the resulting paper score before publishing a score results in substantial robustness to contamination with noise and unreliable reviewers, as done in Figure \ref{fig:4}C. Note that the minimum certainty constraint is required to achieve the performance gains, even when using the oracle with perfect information about reviewer quality (Supplementary~Figure~\ref{figsupp:sf4}). 

In addition, given that our generative model includes two qualitatively distinct types of reviewers (those who issue scores with some correlation to the underlying paper quality and those who issue pure noise), we consider entirely thresholding out (removing from consideration) the bottom percentiles of reviewers instead of weighting their reviews by the Bayesian MSD review quality factor,  reasoning that this bin will contain mostly the uncorrelated reviewers. 

Indeed, if we could directly use a reviewer quality "oracle" estimator, which would have perfect information about the distribution of each reviewer's deviations from CAS, including thresholding out bots, we could achieve a much higher paper quality estimation (Figure \ref{fig:4}C, dashed black line). 
However, in the real-world case where ground truth paper scores are unknown, we find that, surprisingly, thresholding reviewers by their estimated reviewer quality (Figure \ref{fig:4}C, solid red line) does not yield a prediction improvement over the full Bayesian weighting (black line).

The cause of the discrepancy between the oracular review quality thresholding and the normally estimated review quality thresholding result becomes clear when we examine the distributions of estimated reviewer quality, Figure \ref{fig:4}D: even though the unreliable reviewers have a ground-truth reviewer quality of 0, as they become a larger fraction of the reviewing population, their reviews lead other unreliable reviewers to achieve a broad distribution of reviewer quality scores, meaning they are no longer confined to the lower quality bins. Thresholding out the lowest reviewer quality bins does not target exclusively or even a majority of actual low-quality reviewers. This result suggests that it might be possible to further improve the fidelity of review with a ``warm start'' or online reviewing process, in which the platform begins with high-quality reviewers. As new reviewers join, they must work their way up (in the sense of obtaining above-threshold review quality scores from established users) for their reviews to influence paper quality scores and reviewer scores. We investigate this scenario in a subsection below.

\begin{figure}[h!]
\centering
\includegraphics[width=0.7\linewidth]{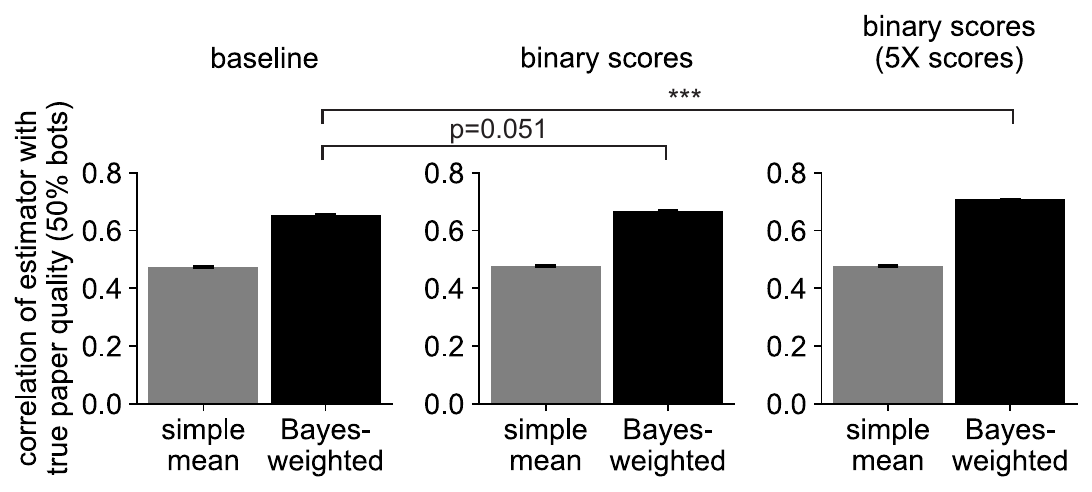}
\caption{{\bf Binarized scoring of reviews does not reduce performance.} 
{\em Left:} The baseline condition, where all scores and reviews are real numbers between 0 and 1. {\em Center:} The "binary scores" condition, where the reviews of reviews (but not paper review scores themselves) are binarized using a threshold of 0.5, does not strongly reduce performance. {\em Right:} The binary scores condition, where the number of binary review of reviews scores are 5 times as numerous as on the left (under the assumption that it's significantly easier to provide a binary label than a real score), leads to improved performance relative to fewer non-binary scores, ***p$<$0.0001.}
\label{fig:7}
\end{figure}

\paragraph{Binary (up/down) scoring of review quality does not substantially degrade performance}
In this section, we investigate the effects of binarizing the scores of others' reviews on the platform ({\em binary scores}). We use the same model from Figure \ref{fig:4}, which runs for 5 ``years." On average, each user writes 3 reviews and provides 10 scores per year (or 50 scores in the 5X scores condition). For this analysis, we fix the proportion of bots on the platform to be $50\%$. We find that there is a borderline but non-significant difference between performance of the Bayes-weighted measure between the baseline and binary scores condition ($P\approx 0.05$), Fig. \ref{fig:7}, suggesting that the platform does not suffer from switching from real number scores in $[0, 1]$ to simpler binary scores which might require less effort from the reviewers. Furthermore, we conjecture that the ease of assigning a binary score will incentivize users to provide more of those scores, improving the overall platform performance; and indeed, we notice a significant increase in the correlation between estimated and real paper quality for the Bayes-weighted metric when increasing the number of scores of others' reviews by the factor of 5 (while keeping the number of reviews per user the same).

\begin{figure}[h!]
\centering
\includegraphics[width=0.5\linewidth]{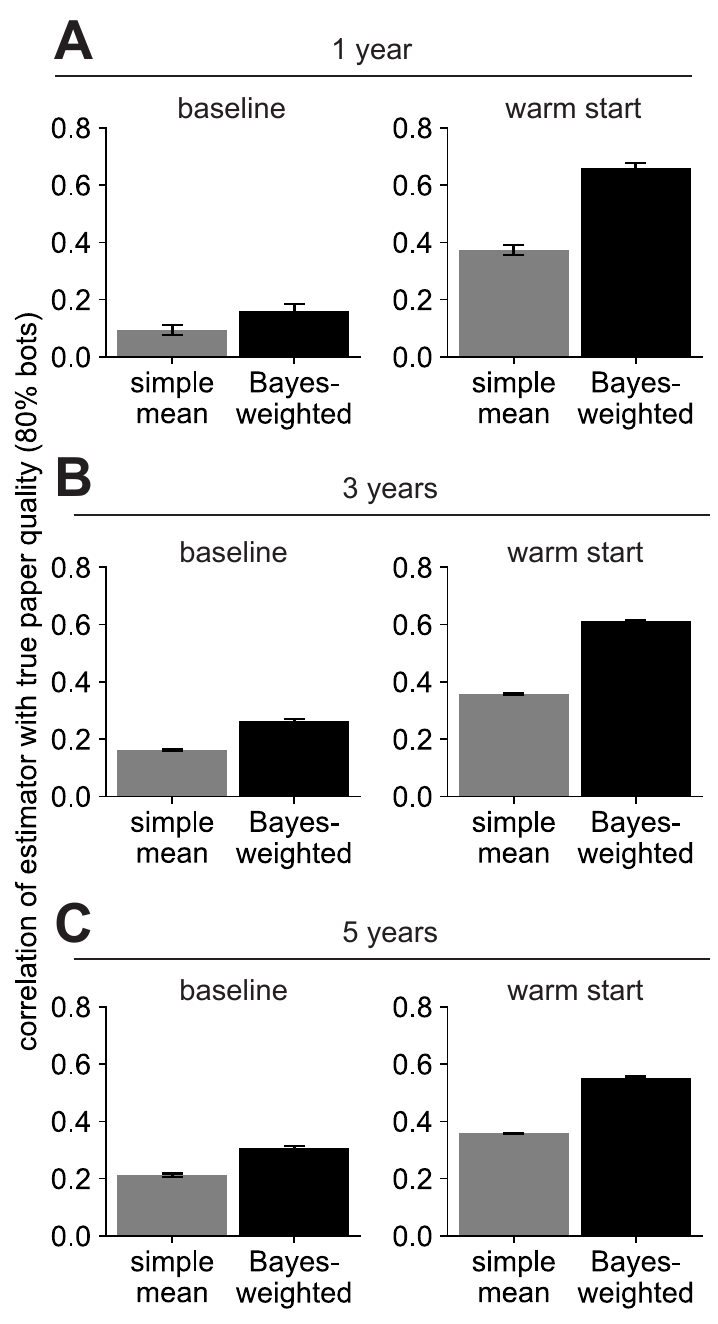}
\caption{{\bf ``Warm start'' platform initialization can result in better performance.} {\em Left:} Performance in the baseline condition, where the initial reviewers are randomly picked from the pool. {\em Right:} Warm start condition: starting from the same pool as in the left, initial reviewers are picked (without replacement) to be the best available reviewers. Subsequent reviewers are selected randomly from the remainder of the pool. Results are for simulations over 1 (A), 3 (B), or 5 (C) years, with 80\% of the users designated as bots. The warm start advantage persists over years but gradually decreases.}
\label{fig:6}
\end{figure}

\paragraph{``Warm Start''} Next, we investigate the effects of initializing the platform with the best available reviewers, a process we will call a {\em warm start}. We consider how this choice might change the long-term dynamics on the platform. We create simulations using the same model as Figure \ref{fig:4}A, running them for multiple ``years". In each year, we simulate the onboarding of new users, content creation, and user exits. Specifically, the simulation begins with 500 users and 20 content items (papers) per user. Each year, 2000 new users join the platform, while 10\% of existing users exit. Users do not generate new content after onboarding, but they do contribute reviews: on average, each user writes 3 reviews and provides 10 scores per year. Here we only quantify an unthresholded warm start (everyone gets a vote, but the platform is seeded with good reviewers), rather than the more aggressive policy of initially assigning zero weights to incoming reviewers until their reviewer quality reaches a threshold. 

To assess the value of a warm start process, we compare two scenarios: one in which the initial 500 users are drawn randomly from the population, and one in which they are selected (without replacement) from the top-performing reviewers (Figure \ref{fig:6}A-C). In both cases, the distribution of review qualities in the user pool is the same; the only difference is the ordering. Results show that warm-starting with high-quality reviewers leads to significantly higher correlations between the estimated and true paper quality, and the effect is largest for the Bayes-weighted measures. The effects are largest in early years and decay gradually over time, but with a persistence of several years. 

\subsection{Extrinsic reviewer incentives} 
There are many reasons why scholars participate in peer review: Altruistic contribution to one's community and the chance to shape the content, claims, citations, and messaging of papers. These intrinsic incentives exist without the community granting overt recognition of the contributions made by reviewing, and without specific incentives delivered to review particular papers. Additionally, in the system outlined above, reviewer quality and thus their impact (weighting) on a paper's assessment is determined based on their quality score, not on the number of reviews they have written (Figure \ref{fig:5}A). Thus, there is an natural bias towards quality over quantity. 

Here we briefly consider additional extrinsic incentives that could further improve reviewer participation, quality, or distribution over papers. 

\paragraph{Reviewer quality publicization}

The computed reviewer quality score can be used to explicitly recognize the quality of a reviewer's work, in the form of a quality certificate. This certificate could be generated publicly for each reviewer without attaching their anonymous reviewer name, thus preserving review anonymity while providing public acknowledgment. Quality certification could be used in various ways, including for promotion or tenure decisions, for public display on websites, and so on. This recognition and accountability would provide additional extrinsic incentives, beyond the intrinsic ones, for writing higher-quality reviews \citep{pascual-ezamaPeerEffectsUnethical2015}. 

\paragraph{Scarcity incentives} Some reviewers may be inclined to write vastly more paper reviews than others, diluting the impact of the rest. We propose explicitly capping the number of reviews than can be written by each user per year, with a possible bonus of 1-2 additional reviews allowed to reviewers depending on their quality score. This makes explicit that quality not quantity determines reviewer scores, and  guards against crowd effects in the review process. In addition, as scarcity creates value, we believe that restricting the ability to review will make reviewing a more valued activity. We encourage future modeling work to take into account human psychology around scarcity to determine the quantitative effects of review caps on reviewer participation and quality. 

\paragraph{Incentive to broaden review coverage across papers}
A common dynamic in nature including human social interactions is the rich-get-richer process, one in which popular or large items become more popular or larger, leading to heavy tails in various distributions. These dynamics are evident in ``viral" content and viral content creators on many social media systems \citep{Narayanan2023-bs}. In the context of bottom-up peer review, a concern is that a few papers might come to garner the lion's share of reviewer attention while many are ignored. Though capping the number of reviews each individual can contribute partially mitigates this effect, it is still possible for a large fraction users to decide to review an already ``viral" paper. 

We consider an information-gain metric to provide further incentives for broad coverage of reviews. Our Bayesian paper evaluation approach process naturally generates an estimation certainty measure per paper (Eq. \ref{cert_est}), which indicates which papers require more reviews. In a preliminary experiment, we assume that the user process of selecting which papers to review is based on the so-called \emph{Chinese restaurant process} (CRP), in which the probability of selecting a paper is proportional to how many reviews the paper already has ($p\propto \#\text{reviews}$, a common dynamic in human group behavior; Figure \ref{fig:5}B, left). With this process, which we treat as the baseline condition, a highly skewed popularity distribution emerges, in which a small minority of papers amass most reviews (Figure \ref{fig:5}B, left). Next, we define a reward signal for each paper, equal to $\Delta \sigma^2_\text{total}$ -- the \emph{amount of reduction} in the uncertainty in the paper's score that would be provided by this reviewer were they to review the paper. We assume that the reward modifies reviewer behavior by convolving the CRP with the reward signal, $p\propto \#\text{reviews}\cdot\Delta \sigma^2_\text{total}$. When reviewer behavior is modulated by the uncertainty-reduction reward in this way, the result is a substantially broader coverage of reviewed papers across the simulated platform (Figure \ref{fig:5}B, right), avoiding the rich-get-richer phenomenon. This suggests that providing reviewers with additional incentives based on the uncertainty-reduction metric could be a powerful tool to obtain a broad reviewer coverage of papers. 

\begin{figure}[h!]
\centering
\includegraphics[width=1.0\linewidth]{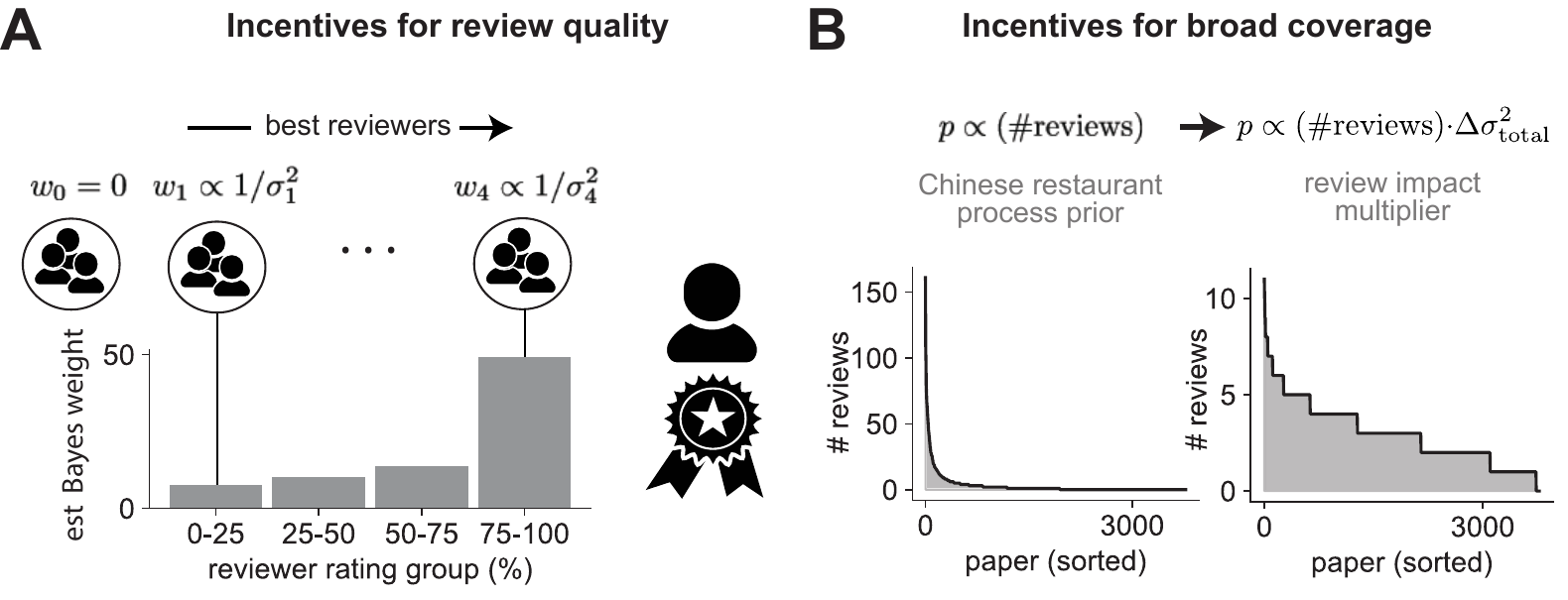}
\caption{{\bf Embedding incentives for writing high-quality reviews and ensuring broad coverage within the open platform framework.} 
(A) {\em Left:} Bayes weighting awards high-quality reviewers with more influence on the final paper score estimation, providing one type of incentive for providing quality reviews. {\em Right:} In addition, users could publicly show and use for promotion and tenure their reviewer ratings, another incentive for high-quality review. 
(B) {\em Left:} Distribution of reviews per paper, assuming reviewers select papers according to a Chinese restaurant process (CRP). {\em Right:} Rewarding reviewers based reducing the uncertainty in the reviewed paper's quality estimate can help flatten the distribution.}
\label{fig:5}
\end{figure}

\section{Discussion}
\paragraph{Summary} 
This paper makes several contributions: we evaluated two real-world peer-review datasets to determine how well reviewer scores correlate with each other, and found very low levels of agreement between reviewers. Using one of the datasets, we showed that there is little correlation between authorship quality and reviewer quality (by computing the community averaged score for paper quality for a subset of papers with a large number of reviews and using it as a metric to assess both authorship quality and reviewer quality per reviewer), so that the authorship quality of an individual is not predictive of their reviewer quality. 

Given these and other quantified challenges in extracting meaningful paper quality scores from small numbers of noisy reviewers, we proposed and assessed Bayesian weighting methods to estimate paper quality based on 1) an empirical estimate of reviewer quality from paper review data or 2) a direct additional evaluation of reviewer quality by other reviewers. In the low number of reviews per paper regime, we showed that Bayesian paper scoring with both proposed reviewer estimation methods significantly outperform standard averaging. In the first method, we mitigated the problem of estimating reviewer quality given the small numbers of reviews written by each individual with a method for binned (collective) estimation of reviewer quality. In addition, we showed that using direct and noisy but unbiased reviewer/reader feedback to determine reviewer quality (method 2) is better than indirect estimation of reviewer quality (method 1) paper quality estimation, even when user feedback has low bandwidth, in the form of binary feedback (simple up- or down-voting). Thus, we have shown through analysis of open review data and modeling that there are simple and effective ways to mitigate the problem of small-$N$ noisy reviews in estimating paper quality, even with a majority of completely unreliable but unbiased agents.
An important direction for future work is to investigate whether combining both reviewer quality estimation methods (the empirical estimate based on deviation from community average scores and the direct peer evaluation of reviews) could yield further improvements in paper quality estimation.

Finally, we explored and quantified the potential of specific metrics as incentives to avoid rich-get-richer snowballing phenomena in paper coverage by reviewers. We showed that an information gain metric could, if implemented as an incentive that influenced reviewer choice, lead to much broader and even coverage of the set of available papers.

\paragraph{Related work} 
Open peer review frameworks for scientific publishing, where reviews can be made public and evaluated, have been widely proposed over the past decade \citep{eisenImplementingPublishThen2020, Ginsparg_1997, Eisen16, Kravitz_2011, Kriegeskorte_2012, Nosek_2012, Stern_2019, Teixeira_da_Silva_2013, LeCun13, 10.3389/fncom.2012.00032}. 
The general idea of users generating commentary and then allowing users to rate the commentary of others has been implemented to considerable success in platforms like StackOverflow and Reddit, with its drawbacks investigated \citep{davisEmotionalConsequencesAttention2021, mazloomzadehReputationGamingStack2024, melnikovDynamicInteractionBasedReputation2018, movshovitz-attiasAnalysisReputationSystem2013, wangReputationStackOverflow2021}. Specifically in the field of scientific peer review, \cite{goldberg2024peerreviewspeerreviews} presented a randomized control trial, carried out at NeurIPS 2022, which showcases the measures of inter-evaluator disagreement, miscalibration, subjectivity, and biases in peer reviews of peer reviews.
Rating user comments is considered a low-bandwidth and simpler exercise than generating commentary. We find that it provides a useful signal for open peer review that simultaneously enables positive incentives as well as more accurate community belief estimation.
Recommendation systems in general and specifically for disseminating scientific research are likewise being investigated \citep{recommendation-systems-15, recommendation-survey-20, achakulvisutScienceConciergeFast2016, koSurveyRecommendationSystems2022, pinedoArZiGoRecommendationSystem2024, putrautamaSCIENTIFICARTICLESRECOMMENDATION2023, shaniEvaluatingRecommendationSystems2011}.
In general settings, crowd-based metrics have been proposed to extract the wisdom of the crowd from the noisy and sometimes on average incorrect signal \citep{prelecSolutionSinglequestionCrowd2017, kamedaInformationAggregationCollective2022, leeUsingCognitiveModels2024, palleyExtractingWisdomCrowds2019}. We believe that the present work, based on concrete suggestions for improving peer assessment and evaluation through generative modeling and analysis of open review data, adds an actionable contribution for the future design and implementation of new systems. 

\paragraph{Outlook: How things could be} The volume of scientific publication has been increasing exponentially, and the difficulty of securing accurate high-quality paper assessments, in a timely way, has grown in tandem \citep{neurips-process-16, kuznetsov2024naturallanguageprocessingpeer}. 

Generating, using, and making available lifetime reviewer scores can incentivize scholars to write reviews and focus on quality, the core challenge for potential open review platforms. We believe this to be the case for two reasons related to innate scholarly drive: the first is a desire for impact (a reviewer's assessment of a paper is weighted more heavily if they have a higher reviewer score), and the second is a desire for recognition (a reviewer's work can is visibly quantified and acknowledged in the form of a score/certification that can be used for promotion and tenure decisions and as a badge of honor). 

The scientific community has shied away from bottom-up peer review to mitigate the potential for low-quality assessments. We conceive of the process as a dynamical system, in which the right mixture of estimation and incenitives can produce self-organized high-quality reviewing ecosystems. Here, we have taken small steps to explore possible metrics and incentives. Implementing and testing these ideas in open publishing platforms should help to provide real-world tests and improvements of such methods, possibly leading to a new process for scientific publishing.

\section*{Acknowledgements}
This work has been supported by NSF-CISE award IIS-2151077 under the Robust Intelligence program and by the Simons Foundation SCGB program 1181110. All authors would like to thank Laurence Hunt, Mariya Toneva, Steven Scholte, and the organizing committee of the 2023 Conference on Cognitive Computational Neuroscience for their support and feedback throughout the project. A.Z. expresses gratitude to Professors James DiCarlo and Rebecca Saxe from the Department of Brain and Cognitive Sciences at MIT for their feedback, guidance and inspiration in earlier stages of this project.

\bibliography{references, references_2, paperpile_cleaned, references_Andrii_latest}

\section*{Code Availability}
All code associated with analyses in this paper will be made available upon publication.

\section*{Appendix 1. MSD Derivations}
\subsection*{MSD of the Simple Mean}
Reviewer \( i \)'s score for paper \( j \) is given by: $s_{ij} = q_j + \epsilon_{ij}$,
where \( \epsilon_{ij} \sim \mathcal{N}(0, \sigma_i^2) \). The simple mean estimator for the paper's quality is:
\[
\hat{q}_j^{\text{simple}} = \frac{1}{n} \sum_{i=1}^{n} s_{ij}.
\]
The mean squared deviation (MSD) is the expected squared difference between the estimator and the true quality:
\[
\text{MSD}(\text{simple mean}) = \mathbb{E}\left[\left( \frac{1}{n} \sum_{i=1}^{n} \epsilon_{ij} \right)^2 \right].
\]
Since \( \epsilon_{ij} \)'s are independent, the variance of the estimator is:
\[
\text{MSD}(\text{simple mean}) = \frac{1}{n^2} \sum_{i=1}^{n} \sigma_i^2
\]

\subsection*{MSD of the Bayes-Optimal Estimator}

The Bayes-optimal estimator weights reviewers by \( w_i \propto 1/\sigma_i^2 \), so:
\[
w_i = \frac{1/\sigma_i^2}{\sum_{k=1}^{n} 1/\sigma_k^2}.
\]
The Bayes-optimal estimator is given by:
\[
\hat{q}_j^{\text{Bayes}} = \sum_{i=1}^{n} w_i s_{ij}.
\]
The MSD is the expected squared difference:
\[
\text{MSD}(\text{Bayes-optimal}) = \mathbb{E}\left[\left( \sum_{i=1}^{n} w_i \epsilon_{ij} \right)^2 \right].
\]
The variance of this estimator is:
\[
\text{MSD}(\text{Bayes-optimal}) = \sum_{i=1}^{n} w_i^2 \sigma_i^2.
\]
Substituting \( w_i = \frac{1/\sigma_i^2}{\sum_{k=1}^{n} 1/\sigma_k^2} \) gives:
\[
\text{MSD}(\text{Bayes-optimal}) = \frac{1}{\sum_{i=1}^{n} 1/\sigma_i^2}.
\]

\section*{Supplementary Materials}

\begin{figure}[h]
\centering
\renewcommand{\figurename}{Supplementary Figure}
\setcounter{figure}{0}
\renewcommand{\thepage}{S\arabic{page}}
\renewcommand{\thesection}{S\arabic{section}}
\renewcommand{\thetable}{S\arabic{table}}
\renewcommand{\thefigure}{S\arabic{figure}}
\includegraphics{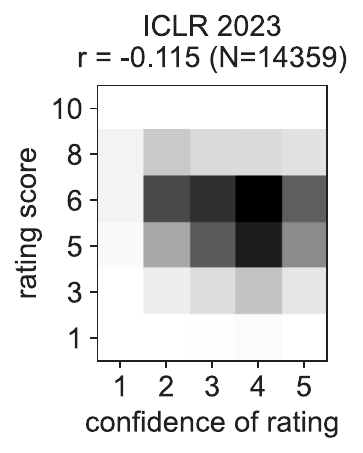}
\caption{Confidence of ratings is inversely correlated to the assigned scores. In the ICLR2023 open peer review data, confidence of assigned ratings was inversely correlated to the assigned scores ($r=-0.115, p<0.001, N=14359$). For every review of every submission in ICLR2023, this plot shows the overall paper score given by the reviewer on the y-axis and the corresponding confidence assigned by the reviewer on the x-axis.}\label{figsupp:sf1}
\end{figure}

\end{document}